\documentclass[10pt, format=manuscript,nonacm]{acmart}

\AtBeginDocument{%
  \providecommand\BibTeX{{%
    \normalfont B\kern-0.5em{\scshape i\kern-0.25em b}\kern-0.8em\TeX}}}

\usepackage{array}
\usepackage{booktabs}
\usepackage{hyperref}
\usepackage{graphicx}
\usepackage{threeparttable}
\usepackage{setspace}
\usepackage[marginal]{footmisc}
\makeatletter
\def\UrlAlphabet{%
  \do\a\do\b\do\c\do\d\do\e\do\f\do\g\do\h\do\i\do\j%
  \do\k\do\l\do\m\do\n\do\o\do\p\do\q\do\r\do\s\do\t%
  \do\u\do\v\do\w\do\x\do\y\do\z\do\A\do\B\do\C\do\D%
  \do\E\do\F\do\G\do\H\do\I\do\J\do\K\do\L\do\M\do\N%
  \do\O\do\P\do\Q\do\R\do\S\do\T\do\U\do\V\do\W\do\X%
  \do\Y\do\Z}
\def\UrlDigits{\do\1\do\2\do\3\do\4\do\5\do\6\do\7\do\8\do\9\do\0}
\g@addto@macro{\UrlBreaks}{\UrlOrds}
\g@addto@macro{\UrlBreaks}{\UrlAlphabet}
\g@addto@macro{\UrlBreaks}{\UrlDigits}

\begin{document}

\title{Data-Driven Decision Making in COVID-19 Response: A Survey}

\author{Shuo Yu}
\affiliation{%
\department{School of Software}
\institution{Dalian University of Technology}
\city{Dalian}
\postcode{116620}
\country{China}}
\email{y\_shuo@outlook.com}

\author{Qing Qing}
\affiliation{%
\department{School of Software}
\institution{Dalian University of Technology}
\city{Dalian}
\postcode{116620}
\country{China}}
\email{QingqingBai@outlook.com}

\author{Chen Zhang}
\affiliation{%
\department{School of Software}
\institution{Dalian University of Technology}
\city{Dalian}
\postcode{116620}
\country{China}}
\email{chen.zhang07@outlook.com}

\author{Ahsan Shehzad}
\affiliation{%
\department{School of Software}
\institution{Dalian University of Technology}
\city{Dalian}
\postcode{116620}
\country{China}}
\email{ahsan.shehzad@outlook.com}

\author{Giles Oatley}
\affiliation{%
\department{School of Engineering, IT and Physical Sciences}
\institution{Federation University Australia}
\city{Ballarat}
\postcode{VIC 3353}
\country{Australia}}
\email{g.oatley}

\author{Feng Xia}
\affiliation{%
\department{School of Engineering, IT and Physical Sciences}
\institution{Federation University Australia}
\city{Ballarat}
\postcode{VIC 3353}
\country{Australia}}
\email{f.xia@ieee.org}

\renewcommand{\shortauthors}{Trovato and Tobin, et al.}

\begin{abstract}
 COVID-19 has spread all over the world, having an enormous effect on our daily life and work. In response to the epidemic, a lot of important decisions need to be taken to save communities and economies worldwide. Data clearly plays a vital role in effective decision making. Data-driven decision making uses data related evidence and insights to guide the decision making process and to verify the plan of action before it is committed. To better handle the epidemic, governments and policy making institutes have investigated abundant data originating from COVID-19. These data include those related to medicine, knowledge, media, etc. Based on these data, many prevention and control policies are made. In this survey paper, we summarize the progress of data-driven decision making in the response to COVID-19, including COVID-19 prevention and control, psychological counselling, financial aid, work resumption, and school re-opening. We also propose some current challenges and open issues in data-driven decision making, including data collection and quality, complex data analysis, and fairness in decision making. This survey paper sheds light on current policy making driven by data, which also provides a feasible direction for further scientific research.
\end{abstract}




\maketitle

\section{Introduction}
\label{sec:introduction}
The COVID-19 pandemic is an ongoing epidemic of the coronavirus disease~\cite{cao2020covid}. The epidemic was declared a public health emergency of international concern by the World Health Organization (WHO) and it is impacting all aspects of life. At the same time, the COVID-19 epidemic is also generating a large amount of data that could be used to effectively guide decision making in the COVID-19 response~\cite{Thomas2021o}. These emergency responses include the detection, prevention and control of the disease, and the recovery from global economic disruption, which is potentially the largest global recession since the Great Depression \cite{verschuur2021observed,schwab2021real}.

Important decisions to be made include postponing or cancellation of sporting, religious, political, cultural activities, and those related to severe stock shortages caused by panic buying \cite{heymann2020covid}. It has been reported that schools, universities, and colleges in more than 180 countries have either closed on a national or local basis, impacting most of the student population worldwide \cite{viner2020school}. Due to the epidemic, vast amounts of data are being generated, which could be used in data-driven decision making to cope with this disease and to respond to relevant emergencies.

Data-driven decision making is the process of using evidences and insights derived from data to guide the decision making process and to verify a plan of actions before it is committed~\cite{cippa2021data,zhang2020data}. It plays a vital role in the success of businesses, industries, governments, and even an individual's life. We continually make decisions during every aspect of our lives~\cite{nisioti2021data,catelani2020optimizing}. Indeed, success is based on these decisions, with correct decisions helping to achieve goals meanwhile wrong decisions resulting in failure. Data is the key element, playing the most significant role in effective decision making. It is the primary input for many fields of science like data mining, big data, machine learning, and data science~\cite{liu2019data,bedru2020big,liu2019shifu2}. The objective of all these fields is to analyse data and to provide insights that actually assist to perform effective decision making. Data-driven decision making is changing the field in all professional endeavours, including business, medicine, and education~\cite{dolgin2020core}. Here, experts gather and analyse data to effectively prompt their decision making~\cite{Leung2020,Gheluwe9239038}. There has been broad adoption in industry, with businesses using data analytics at every stage in their operation, from the management of supply chains to the creation of competitive advantage.
\begin{figure*}[!t]
  \centering
  \includegraphics[width=0.8\textwidth]{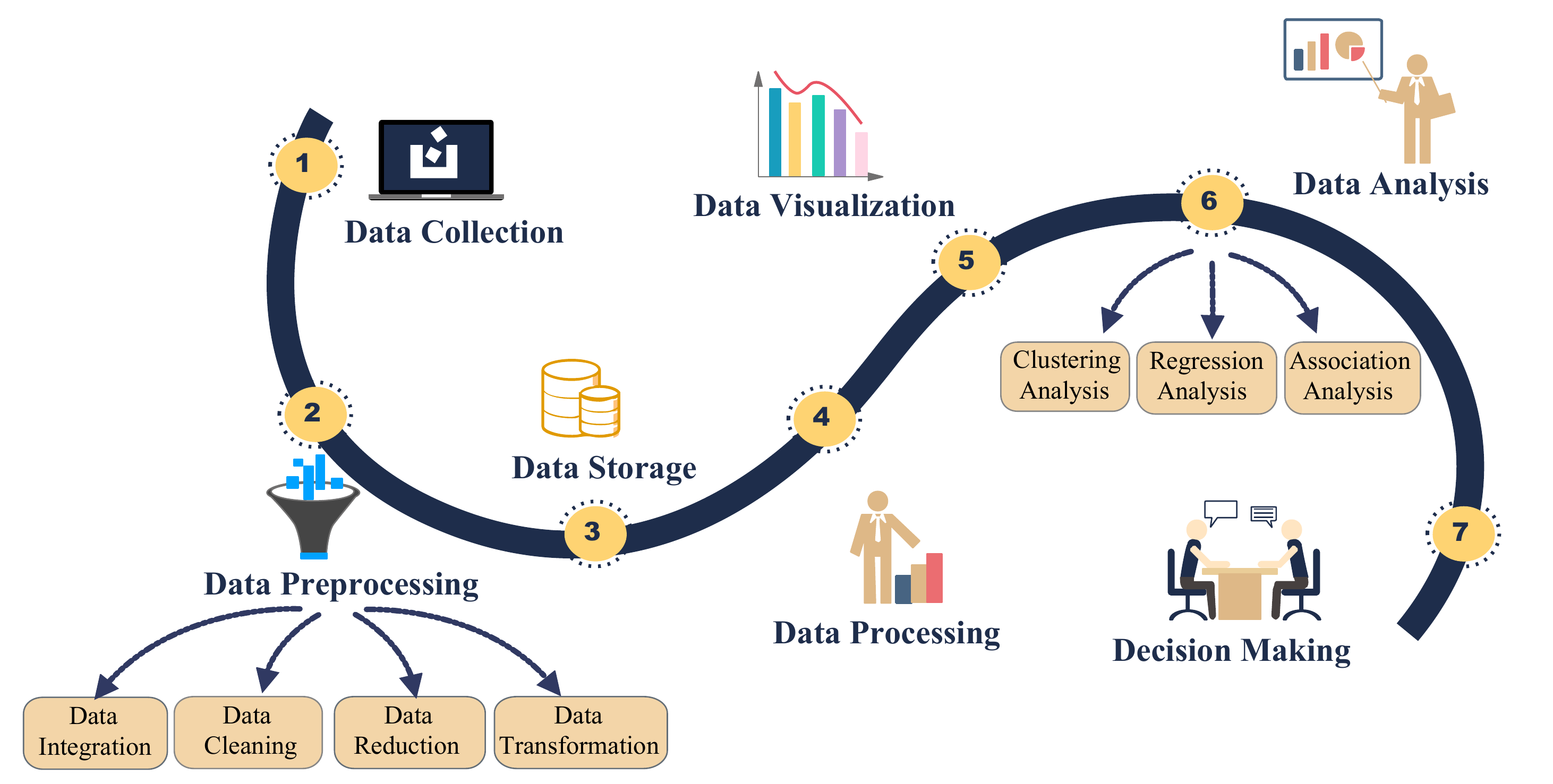}
  \caption{Big data processing stages.}
  \label{data-decision}
\end{figure*}

Data play a vital role in effective decision making \cite{ellul2015big}. In the response to the COVID-19 epidemic, a lot of important decisions need to be taken to save communities and economies worldwide. In this paper, we review the different decisions made in response to the COVID-19 epidemic from the perspective of data and subsequent insights. This includes the use of data for the epidemic prevention and control, psychological counselling, the protection of livelihoods, and the financial aid efforts to recover the economy~\cite{Xin9209939}. This survey also includes the data-driven decisions made related to resumption of work and schools. We also describe the challenges and open issues which hinder the development of data-driven decisions based on insights and analytics, and these challenges will open new directions for future research.

The paper is organized as follows. The literature and techniques of prevention and control are introduced in Section~\ref{sec:2}. The Psychological effects of the disease and related counselling is described in Section~\ref{sec:3}. The needs for financial aid for the protection of lives and recovery of  economy is described in Section~\ref{sec:4}. Data-driven decisions in the return to workplaces and schools are described in Section~\ref{sec:5} and Section~\ref{sec:6}, respectively. The key challenges and issues for data-driven decision making specifically related to the COVID-19 epidemic are explained in Section~\ref{sec:7}. Section~\ref{sec:8} concludes the paper.

\section{COVID-19 Prevention and Control}
\label{sec:2}
\begin{table*}[!t]
  
  \centering
  \caption{COVID-19 Dataset}
  \label{tabdatasets}
  \resizebox{1.0\linewidth}{!}{   
    \begin{threeparttable}
      \begin{spacing}{1.19}
        \begin{tabular}{p{3cm}p{4.5cm}p{10cm}p{3.3cm}}
          \toprule[2pt] 
          \textbf {Category} & \textbf {Title} &\textbf {Main Content} & \textbf {Provider}  \\
          \midrule[1pt]
          Epidemic & Worldometer COVID-19 Data & Confirmed cases or deaths in different countries and regions & Worldometer \\
          Scientific Research & COVID-19 Open Research Dataset (CORD-19) & More than 45,000 academic articles are freely available, including 33,000 full texts on COVID-19 and coronavirus family viruses & Allen Institute for AI\\
          Knowledge Graph & COKG-19 & The COKG-19 contains 505 concepts, 393 attributes, 26,282 entities and 32,352 knowledge triples, covering medical treatment, health, materials, prevention and control, scientific research and people, etc & AMiner \\
          Migration & Baidu migration & Population migration during the Spring Festival in China & Baidu Map\\
          Social media & COVID-19-TweetIDs & The repository contains an ongoing collection of tweets IDs associated with the novel coronavirus COVID-19, which commenced on January 28, 2020  & Github (Emily Chen)\\
          Medical & COVID-chestxray-dataset & It an open database of COVID-19 cases with chest X-ray or CT images & Github (Joseph Paul Cohen)\\
          Traffic & TSA: Airport confirmed case data & American airports have tested for confirmed cases in the past 14 days & Transportation Security Administration\\
          \bottomrule[2pt]
        \end{tabular}%
      \end{spacing} %
    \end{threeparttable}      
  }
\end{table*}%

The global spread of COVID-19, which has affected hundreds of countries and regions, is a huge challenge to the ability of governments to manage crises~\cite{de2020initial}. Early response and timely action are extremely important in the fight against COVID-19. Hern{\'a}ndez-Orallo et al.~\cite{hernandez2020evaluating} have proven that using data analysis to track the disease source, isolating the infected person in time, and controlling social distance are effective methods in preventing further spread of COVID-19. There are some studies that illustrate the response of different countries in the initial period of the outbreak and the later control measures~\cite{kraemer2020effect,atk2020G20,tomar2020prediction}. These works show that despite different countries having specific medical conditions, there are still several decisions that can be made to cope with disease outbreaks. If certain measures of prevention and control are adopted according to the situation in the early stages, the spread of the epidemic can be slowed down. The development of science and technology has made it possible for us to mine knowledge from big data, which can guide decision making in epidemic prevention and control as well as policy formulation~\cite{anastassopoulou2020data}.

\subsection{Data-Driven Decisions}

With the continuous spread of the epidemic, many types of information and data related to the epidemic are continually updated on the Internet. There is a huge amount of information, including global epidemic data, COVID-19 related research statistics, medical equipment details, transportation information and so on. The integration and classification of these forms of information are important for epidemic prevention and control. We have categorized some commonly-used datasets by information type, as shown in Table~\ref{tabdatasets}. By using this, researchers can select appropriate datasets for analysis according to their needs.

To make full use of the data, it is necessary to process the data. The big data processing stages include data collection, data pre-processing, data storage, data analysis, data visualization, data application, etc, which are shown in Fig.~\ref{data-decision}. Through data analysis, technicians can effectively process raw data and aggregate related information. Subsequently, valuable information can be derived and then  used to interpret and predict the development of the epidemic. Based on more comprehensive information or stronger evidences, decision-makers can make more scientific decisions~\cite{lytras2017big}.

Policies are generally made dynamically. This is because policies need to be adjusted in time according to specific situations. In modern society, decisions are no longer made based on empirical judgements, but are generally made based on big data~\cite{mcafee2012big}. Big data statistics are used in order to permit policy makers to rapidly assess the epidemic risk. With respect to COVID-19, related decisions and relevant plans are also adjusted correspondingly based on big data analysis~\cite{power2014using}. Big data technologies provide accurate information for the decision making and analysis of the headquarters for the epidemic prevention. With the help of  big data technologies, epidemic prevention and control can be carried out in an orderly manner, so that the epidemic can be effectively controlled within a determined period of time.

\subsection{Tracing Close Contacts}
Close contact tracing is a common intervention to control outbreaks of infectious diseases~\cite{hellewell2020feasibility}. In the early stages of the epidemic, traditional door-to-door surveys and paper form filling were still adopted in most areas, which was organized and implemented step by step~\cite{li2020early}. If the execution was not strong enough or the provided information was wrong, it would affect the judgement and prevention of the epidemic. After that, each person's action trajectory was generated by analyzing people's location information. The responsible persons were isolated and additionally the close contacts, who were identified based on the action trajectory and reported information. If some of these people are found to have abnormal health, the authorities can provide them with timely help~\cite{fisher2020global}.

The ``Big data and Grid" method is adapted to improve the screening efficiency in inspecting the general trend of people mobility~\cite{li2020early,xia2018commag,xia2018tii}. Additionally, the government release information about ``epidemic community", ``infection routes" and ``patients on the same journey". People can see if there are new cases within their community on the web via HealthMap\footnote{\url{https://www.healthmap.org/covid-19/}} at any time. The travel information of suspected infected persons can also be queried, to quickly determine whether they are close contacts. To some extent, this also reduces for the relevant departments the pressure of screening. National Health Insurance data and the database from the Immigration Department can be merged for better decision making. After analyzing the data, medical staff can make judgements based on the patient's travel records and clinical symptoms~\cite{wang2020res}. Smart card data can also be employed to identify suspected patients and isolate them at an early stage, which will effectively reduce the spread of the epidemic.

Smartphone solutions have also been developed to enhance tracing close contacts. Location-based technologies are employed in a range of applications that utilize location sensors and capabilities of smartphones~\cite{9117157,xia2014exploiting,xia2014community}. Proximity details, identity, location data, as well as other condition information can be captured by apps on the smartphone. Apps such as TraceTogether employ a data-driven community-based approach to share position data and time stamps. Meanwhile, to protect privacy of users, TraceTogether will overwrite its data after it collects newer data every 21 days. This is to say, this app only saves position and time data for 21 days. We can see the list of corresponding apps shown in Table~\ref{tab:app}. However, some non-specialized apps such as QR-codes and electronic cards in WeChat and Alipay are not included in the Table.

Tracing methods can be further enhanced by optimizing testing and tracing coverage. More mobile application technologies need to be developed to improve contact tracing effectiveness. Bradshaw et al.~\cite{bradshaw2021bidirectional} found that contact tracing plays a critical role in controlling and preventing COVID-19, but most tracing protocols rely on forward-trace instead of bidirectional tracing protocols. They utilized the bidirectional tracing method, resulting in significant improvement in COVID-19 control. Ng et al.~\cite{9373368} presented a smart contact tracing system named SCT, which employs a smartphone's Bluetooth low energy signals together with machine learning classifiers to detect potential contacts to confirmed cases. The privacy problem should also be considered when releasing the confirmed cases. Publicizing close contact data can help the public better avoid causing COVID-19 infections, but also increases the potential for data breach problems at the same time. The technological limitations and the balance between privacy and tracing effectiveness needs to be taken into full consideration.
\begin{table*}[!t]
  \centering
  \caption{Smartphone applications}
  \label{tab:app}
  \resizebox{1.0\linewidth}{!}{   
    \begin{threeparttable}
      \begin{spacing}{1.19}
        \begin{tabular}{p{1.5cm}p{2 cm}p{1.5cm}p{2cm}p{3cm}p{3cm}p{9cm}}
          \toprule[2pt] 
          \textbf {Number} & \textbf {App Aame} &\textbf {Country} & \textbf {Operating System} & \textbf {Developer} & \textbf {Size} & \textbf {Official Website}\\
          \midrule[1pt]
          1 & Aarogya Setu & India & iOS/Android & National Information Center & iOS:37.1M Android:4.2M & \url{https://www.aarogyasetu.gov.in/}\\
          2 & PrivateKit & USA & iOS/Android & Massachusetts Institute of Technology & iOS:34M Android:8.4M & \url{http://privatekit.mit.edu/}
          \\
          3 & Rakning C-19 & Iceland & iOS/Android &  Icelandic Government & iOS:20M Android:28M & \url{https://www.covid.is/app/is}
          \\
          4 & Stopp Corona & Austria & iOS/Android & Austrian Red Cross & iOS:15.6M Android:6.2M & \url{https://www.stopp-corona.at/}
          \\
          5 & TousAntiCovid & France & iOS/Android & Government Technical Institutions & iOS:76.1M Android:23M & \url{https://www.frandroid.com/telecharger/apps/stopcovid-france}
          \\
          6 & TraceTogether & Singapore & iOS/Android & Government Technical Institutions & iOS:71.4M Android:26.1M & \url{https://www.tracetogether.gov.sg/}
          \\
          \bottomrule[2pt]
        \end{tabular}%
      \end{spacing} %
    \end{threeparttable}    
  }
\end{table*}

To better solve this problem, Aarogya Setu uses contact information to track details. If one of the contacts is confirmed to be COVID-19 positive, then the user is notified to implement aggressive medical interventions. Rakning C-19 runs in the background and stores GPS positions many times per hour for 14 days. These position data will only be saved in the user's phone and no-one else has the access to these data. Stopp Corona transfers data with encryption and anonymous processing. TousAntiCovid requires the user to have an open Bluetooth connection, providing anonymous record services. If a certain user is infected, TousAntiCovid will inform this user and close contacts based on anonymous records.

\subsection{Trend Forecasting}
Forecasting the  trend of development of the epidemic is great significant to the prevention and control of the situation, the distribution of medical resources, economic development, and the arrangement of production activities~\cite{li2020trend,liu2020cri}. By releasing significant amounts of data and statistics, policy makers and the public can understand the development of the epidemic and the effectiveness of its control. For policy makers, if the scale or deterioration of the epidemic  is known as early as possible, they can take corresponding measures to disrupt the chain of transmission and more reasonably allocate medical resources, which can save lives~\cite{fanelli2020analysis}.

Many researchers have proposed different propagation prediction models in combination with their own research directions. Zhou et al.~\cite{9338466} used Autoregressive Integrated Moving Average (ARIMA) model, logistic regression, Susceptible Infective Removed (SIR) model, and improved Susceptible Exposed Infective Removed (SEIR) model to predict the global pandemic. The benefits and drawbacks of all these models are respectively discussed. The global cumulative number of confirmed, cured, and cases of death of both COVID-19 and SARS are employed to better predict the global pandemic trend. The theoretical basis is also provided for prevention and control policies. To accurately forecast the development trend of COVID-19, Kumar and Susan~\cite{9225319} used temporal data from cumulative cases of the 10 most affected countries. Based on ARIMA and Prophet time series forecasting models, the evolution of the COVID-19 outbreak is then modeled and evaluated by strict mathematical metrics such as mean absolute error. It is verified that ARIMA is more effective in forecasting COVID-19 prevalence. The forecasting results will provide potential assistance for governments when planning policies for containing the spread of COVID-19.

\subsection{Hospital Relevant Policies}
COVID-19 has extremely challenged medical processes such as hospital admissions. Data-driven hospital decision making has become more and more important. To better mitigate the epidemic and meanwhile limit collateral economic damage, proper hospital policies should be undertaken. Duque et al.~\cite{Duque19873} presented a strategy to minimize the number of days of disruption. When hospital admissions exceed a certain threshold, short-term in-place placement orders can be triggered. Random optimization is employed to export the triggers to make sure that the hospital admission capacity will not be insufficient. Based on COVID-19 hospitalization data, this work provides a flexible framework, which allows optimization using a relatively small discrete grid. Similar epidemics can be simulated with this proposed model.

The medical examination process can also be improved by using data analysis. Cohort-level clinical data, patient-level hospital data, and census-level epidemiological data can be analysed to develop an integrated optimization model. In the city of Daegu, the South Korean government reorganized the health system along with hospital-level interventions. Equipment are concentrated to relieve the shortage of medical resources. This policy ultimately protected patients and health care staff. The Iranian government added 3,000 hospital beds to the Iran Mall Exhibition based on data analysis. Meanwhile, a 2000-bed army hospital has been opened to accommodate more patients.

While optimizing the medical configuration and treating confirmed cases, scientists from all over the world are working towards effective vaccines for COVID-19. The United States, Australia, Russia, UK, China, South Africa, Korea, and many other countries have made progress towards the COVID-19 vaccine. With the progress of vaccine development, many hospitals have provided vaccine injections.

\begin{figure*}[!t]
  \begin{center}
    \includegraphics[width=0.8\textwidth]{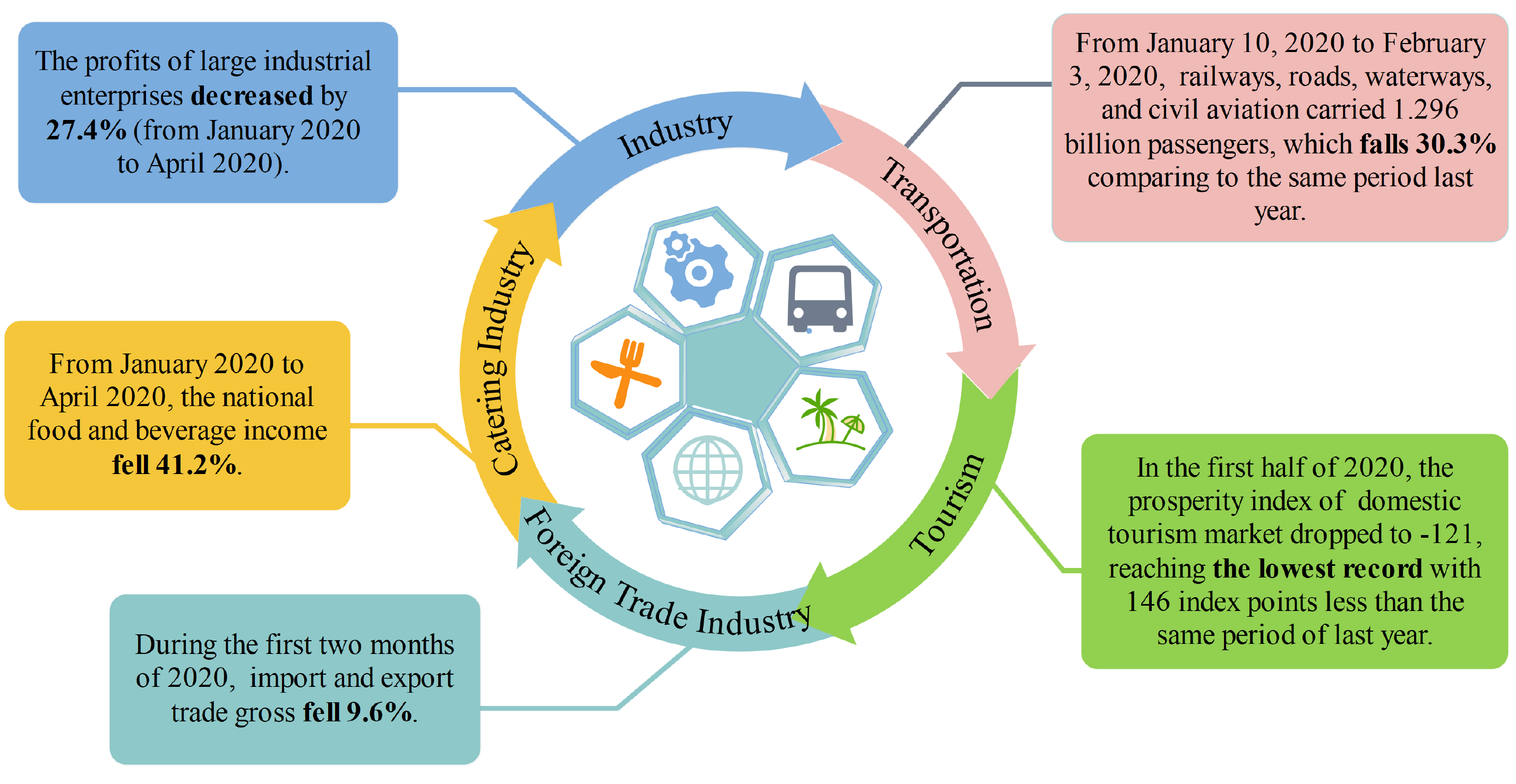}
    \caption{Impact of COVID-19 (Data source: National Bureau of Statistics of China, World Tourism Alliance, Ministry of Transport of China, and General Administration of Customs of China)}
    \label{industry}
  \end{center}
\end{figure*}

\subsection{Influence of the Pandemic}

The outbreak and rapid spread of COVID-19 has caused an enormous impact on every country's economy in a short period of time. The impact on key industries such as accommodation and catering, transportation, tourism, and foreign trade is significant. To prevent the spread of the epidemic, many cities have closed roads, airports, and high-speed trains. Major tourist attractions have been closed and major cultural events cancelled. There has been a sharp decrease in all kinds of dinner parties and a large number of restaurants have been closed, with the stock of prepared dishes being sold at low prices. Workers cannot work normally, and companies have no way to produce and deliver as planned, so the supply chain is interrupted, and many factories face the crisis of closure~\cite{baldwin2020thinking}. The specific impact of the epidemic on these five industries is shown in Fig.~\ref{industry}.

Therefore, taking measures as early as possible to control the epidemic is key to reducing the impact of the epidemic on multiple industries. At present, the epidemic situation has been effectively controlled in many countries.

\section{Psychological Counselling}
\label{sec:3}
The outbreak of COVID-19 has made people face many changes in lifestyle and environment~\cite{kong2018human}. Many places have adopted mandatory isolation measures to prevent the further spread of the virus. Isolation increases the possibility of depression and even makes people ``desperate"~\cite{horton2020offline}. Moreover, the negative reports about COVID-19 illnesses are constantly updated every day. The number of dead cases is increasing, and the number of confirmed cases is increasing daily, resulting in incalculable loss. These series of problems increase people's mental stress and negative psychological states. The large-scale pandemic disease causes panic in people, worrying about the possibility of becoming sick, and meanwhile increasing the fear of death, helplessness, irrespective of being a patient~\cite{North2013}. These cases involve the pain of treatment, the risk of relapse of mental diseases, and several uncontrollable behaviours.
\begin{table*}[!t]
  \begin{threeparttable}
    \caption{Some Decision Making Issued by International Organization}
    \centering
    \label{table_example}
    \begin{tabular}{p{1.5cm}p{4cm}p{2cm}p{9cm}}
      \toprule[2pt] 
      \textbf {Number\tnote{*}} & \textbf {Institution}&\textbf {Date issued} & \textbf {Decision making} \\
      \midrule[1pt]
      
      1&World Health Organization (WHO)&Mar 18, 2020& Mental health and psychosocial considerations during the COVID-19 outbreak\\
      2&Centers for Disease Control and Prevention (CDC) &Apr 30, 2020& Coping with Stress\\
      3&World Federation for Mental Health (WFMH) & Apr 22, 2020& Appeal for National Plans for Mental Health during the Coronavirus Global Emergency\\
      4&National Alliance on Mental Illness (NAMI) & May 12, 2020& COVID-19 Resource and Information Guide\\
      5&Office for the Coordination of Humanitarian Affairs (OCHA) &Mar 17, 2020& Interim Briefing Note Addressing Mental Health and Psychosocial Aspects of COVID-19 Outbreak\\
      6&Bureau of Disease Control and Prevention & Mar 18, 2020& Notice on Issuing the Work Plan for Psychological Counseling of New Crown Pneumonia Epidemic\\
      7&Bureau of Disease Control and Prevention &Feb 7, 2020& Guidelines for the Work of Psychological Aid Hotlines\\
      8&National Health Commission &Feb 15, 2020& Implement Several Measures to Improve Medical Staff's Working Conditions and Physical and Mental Health\\
      9&Ministry of Education &Feb 16, 2020& Suggestions on Guiding Children to Study and Live at Home during Epidemic Prevention and Control\\
      10&National Food Safety Risk Assessment Center & Mar 11, 2020&Advice on Public Mental Health Guidance during New Crown Pneumonia Epidemic\\
      \bottomrule[2pt]
    \end{tabular}
    \begin{tablenotes}
      \footnotesize
      \item[*]Number does not represent the ranking order of the institutions.\\
%
    \end{tablenotes}
  \end{threeparttable}
\end{table*}

More than half of the people were seriously affected by the outbreak of the disease. According to reports, during the SARS outbreak, there were many mental health diseases, such as depression, anxiety, panic, and additionally people developed suicidal tendencies~\cite{ben2004traumatic}. Among patients, 59\% of those hospitalized suffer from the impact of depression and trauma. The importance of psychological protection for medical staff should also be emphasized. What is more, with the expansion of the epidemic, the economic downturn, unemployment, and financial difficulties also affect people's mental health, and its serious results are immeasurable. Therefore, during the prevalence of COVID-19, besides the need for active and effective treatment measures, mental and psychological health guidance is also indispensable.

\subsection{Official Policies}

The WHO issued customized mental health guidance for individuals, which aims to immediately consider people with mental health problems~\cite{world2020mental}. Similar guidelines have also been issued. A large-scale survey which covers 58 countries and more than 100,000 participants was conducted to study the public reaction of government policies\footnote{\url{https://covid19-survey.org/}}. Cross-country data analysis shows that strong government policies can reduce public worries and depression. Moreover, it is also suggested that policymakers should also consider how their decisions affect the mental health of their population. Some international health organizations have issued suggestions on mental health problems during the COVID-19 epidemic, as shown in Table~\ref{table_example}.

 Guidelines for Emergency Psychological Crisis Intervention in Pneumonia Infected in COVID-19 has been issued. Such guidelines put forward psychological intervention points for confirmed patients, suspected patients, medical care, and related personnel such as those who are in close contact with patients (family members, friends, colleagues, etc.), those who do not want to see a doctor publicly, and vulnerable people from other different groups. Li et al.~\cite{li2020progression} collated the documents of specific intervention and guidance measures for different groups in the severe outbreak period. However, with the continuous changes during the epidemic situation, the corresponding psychological guidance measures should be improved for different populations. Duan and Zhu~\cite{duan2020psychological} pointed out the shortcomings of the existing system and proposed corresponding improvement measures for specific problems. The accurate publication of the government's daily disease data-although making people feel panicked  about the increase in number of cases-reduces people's fear of the unknown because the transparency of this information is also crucial to reduce the spread of rumours.

Policy documents from the Interagency Standing Committee entitled ``Addressing Mental Health and Psychosocial Aspects of the COVID-19 Outbreak" have also been released. Wherein, this document summarizes the global definitions of mental health and psychosocial support and introduces mental health and psychosocial responses to COVID-19. The British Psychological Society released information handouts entitled ``Talking to Children about Coronavirus", aiming to provide parents with better guidance when talking about COVID-19 with children. The University of Reading and University of Oxford also gives advice to parents about how to reduce anxiety of children or young people about COVID-19.

Psychological issues during the epidemic can be caused by various reasons, including social distancing, the worldwide lockdown creating economic recession, social boycott, and discrimination, etc. Medical healthcare professionals and staff also suffer from stress, anxiety, and pressure, which also potentially cause psychological issues. To better offer psychological support and advice to front-line medical staff and the public, many organizations and institutions provided remote help. Oxford Centre for Anxiety Disorders and Trauma offers remote treatment of Post Traumatic Stress Disorder (PTSD) and social anxiety with cognitive therapy. Panic disorder can also be helped through remote delivery or teleworking. The British Psychological Society, Division of Clinical Psychology, Digital Healthcare Sub-Committee also provide therapy via video. A variety of websites have been developed to provide support as well. Examples include Support The Workers, COVID Trauma Response Working Group, Intensive Care Society: Wellbeing Resource Library, Coping With Coronavirus, Maintaining health and wellbeing during the COVID-19 pandemic, Psychosocial responses to COVID-19, Second Victim, Just Listening, American Psychological Association-Pandemics, etc.

\subsection{Enterprise Policies}
Significantly, most psychological intervention guiding measures have been taken up by hospitals and psychological counselling agencies. The decreed isolation policies also prevent enterprises from resuming work. Whether it is a private enterprise or a state-owned enterprise, employees are also isolated to their homes like most people. Due to the isolation policies, face-to-face communication is not possible, so hospitals and consulting agencies have jointly built many online consultation platforms~\cite {liu2020online}. Zhang et al.~\cite{Zhang2020} proposed a mental health intervention model, which uses an Internet technology platform to provide timely psychological counselling to patients and their families. Doctors need the courage to overcome disease meanwhile, they need to deal with certain extreme emotions of patients due to illness, who can be physically and mentally exhausted. Therefore, their psychological guidance is also a top priority. Chen et al.~\cite{chen2020mental} detailed the psychological guide measures for medical staff, including the establishment of a psychological intervention medical team, the establishment of a psychological assistance hotline team and psychological skills training.

\subsection{Education Policies}
To cope with the disease, some countries has explicitly closed schools. However, long-term suspension of classes at home will certainly have a serious of negative impacts on students' psychologies \cite{Brooks2020}. Children are more likely to feel pressure after isolation, which is four times higher than those who are not isolated \cite{Sprang2013Posttraumatic}. Research shows that people with higher education are more likely to feel uncomfortable because of their higher awareness of health \cite{Tessa2018Factors}. Cao et al. \cite{CAO2020112934} analysed the psychological status of college students during the new outbreak by questionnaires and found that they all had different degrees of anxiety.

In addition to researching effective measures for the treatment of COVID-19~\cite{jrfm13020036}, many universities also provide online one-to-one psychological counselling services. A public welfare project: ``Combating Epidemic Disease and Psychological Assistance" has been carried out by the jointly organization of many institutions. This provides free psychological assistance to the public and face-to-face psychological counselling services \cite{Tsinghua2020}. Shaanxi Normal University has also published the first national ``Anti-epidemic Psychological Guidance Manual", which is published in paper, electronic versions and audiobook formats. Some university professors even spontaneously organized and recorded videos of exercise at home, which can effectively regulate the physical and mental health of people at home, and help maintain social stability.

\subsection{Technological Innovation}
During the SARS outbreak, because the development of the Internet was not yet mature, and smartphones were not as popular as today. Online psychological counselling services  were not available \cite{chan2007improving}. Today, with the prevalence of mobile phones, the development of Internet services, and the emergence of the 5G era, major medical platforms have provided online psychological counselling services, allowing people to conduct psychological counselling even if they are isolated at home. For example, the Structured Letter Therapy method \cite{articleXiao} and Health Intervention Model \cite{Zhang2020}. Liu et al. \cite{liu2020online} discussed the specific online service counselling during this period. Through psychological questionnaires and counselling, doctors can better understand one's tendencies and guide them in a timely manner through their psychological disorders. According to research, Artificial Intelligence (AI) technology \cite{liu2018artificial,xia2019random} can also help identify people with suicidal tendencies, which also provided a strong assistance for professionals supporting mental health \cite{Just2017}. The WHO and the CDC, to help people cope with their emotions arising from the epidemic of COVID-19, haved employed chat robots to communicate with people and provide emotional communication. Miner et al. \cite{Miner2020} also analysed the advantages and disadvantages of chat robots in mental health guidance, indicating future prevention work.

In general, with the gradual stabilization of the domestic epidemic situation, the psychological health guidance for the public also needs to be given priority consideration. Although government agencies have issued many related guidance programs and achieved initial results, there are still some deficiencies. At present, people are still facing many problems such as long-term shutdown, which makes many enterprises and individuals face the risk of bankruptcy and unemployment. All industries, including education, economy, and tourism, are bearing the consequences of the loss brought about by COVID-19. Domestic epidemic control is relatively stable, but foreign situations are still relatively severe. The most serious problem is that, according to the current research situation, the prevention and control of COVID-19 will be a protracted effort, which will coexist with human beings for a long time. The existence of these problems makes people feel unstable and originates countless psychological problems. Therefore, psychological health guidance is essential.

\section{Financial Aid}
\label{sec:4}
The WHO and the International Monetary Fund (IMF) have never been more important in supporting the global emergency response. Funding needs to be made available in tracking the spread of the virus, ensuring patients receive assistance and primary staff receive essential supplies and information \cite{wang2020combating}. All these decisions about financial support are based on the data of the affected communities and economies. The Governments, agencies, industries, markets, and individuals all have come together to fight the COVID-19 pandemic and to help respond to this global outbreak.

\subsection{Financial Aid for the Protection of Lives and Livelihood}
The world is facing an immense challenge, with the rising pandemic of COVID-19 impacting populations and economies everywhere. A tremendous amount of financial assistance is required to support the communities and economies of the world. Financial assistance should be provided on two priorities: 1) the protection of lives; and, 2) the protection of livelihoods. The protection of life means that countries should put health spending at the top of the priority list. This includes funding health systems receiving monies for doctors, nurses, and hospitals, purchasing medical equipment, and helping the most in need \cite{world2020considerations}. The protection of livelihoods means providing financial support for the provision of lifelines for households and companies during this time of economic problems. This includes cash incentives, wage increases, and tax cuts, helping people meet their needs and helping companies keep operating. For those who have been laid off, unemployment compensation may be temporarily improved by extending its length, increasing the benefits, or relaxing the requirements for eligibility. As far as monetary policy is concerned, providing sufficient liquidity to banks and non-bank finance companies, particularly those lending to small and medium-sized enterprises that may be less able to withstand severe disruption, is crucial at this point \cite{Mckibbin2020}. It is also reported that United Nations International Children's Emergency Fund (UNICEF) has launched an emergency response that will help provide food for children in the United Kingdom who are affected by COVID-19. It is reported that UNICEF will provide grants to 30 local organizations. For example, one of the organizations directly provided 18,000 nutritious breakfasts to 25 schools\footnote{\url{https://news.sky.com/story/covid-19-for-the-first-time-in-its-history-unicef-will-help-feed-kids-in-the-uk-12163515}}.

Though COVID-19 began as a health crisis, it has triggered a grave and unfolding economic crisis with unpredictable loss. People on low-income and those who barely survive on precarious livelihoods have suffered the hardest. According to the United Nations statistics, over 300 million children who rely on school meals might now be at risk of acute hunger. Additionally, social distancing might not work in urban slums and rural households since toilets are shared by multiple families. To make real-time financial response under such circumstances, we need not just appropriate isolation policies, but also normal lives. It is suggested that countries should let people continue their normal lives, that is, to work, earn money, and feed their families. Shop keepers should open and provide services with effective protection such as gloves and masks. Widespread and strict lockdowns might unintentionally lead to more deaths, which affects more the poor community. Otherwise, the global financial institutions will have to write off debts from low-income countries and provide enough resources for recovering economies~\cite{2020Has}.

\subsection{Financial Aid for Economic Recovery}
The global economic crisis triggered by COVID-19 has never  before been seen in history, and there is significant concern as to its effect on people's lives and livelihoods \cite{bethune2020covid}. Most of the situations depends on the epidemiology of the virus and the effectiveness of control steps, all of which are hard to anticipate. In addition, several countries are now facing compounded crises - a health crisis, a financial crisis, and a fall in stock prices, all that interact in complex ways. The estimated loss to global
Gross Domestic Product (GDP) between 2020 and 2021 from the pandemic may be about \$9 trillion, more than the economies of Japan and Germany combined\footnote{\url{https://openknowledge.worldbank.org/handle/10986/33488}}.

A large amount of financial assistance is needed for the recovery of the economy worldwide. When the economic crisis is over, countries will face high rates of debt, bankruptcies, unemployment, and falling wages \cite{nicola2020socio}. The pace of economic recovery will depend on the policies pursued during this crisis. When policies ensure that employees will not lose their jobs, tenants and homeowners are not disadvantaged, businesses prevent bankruptcies, and market and trade networks are preserved, then recovery can take place faster and more smoothly \cite{fernandes2020economic}. In the case of countries that do not have a fiscal environment to implement these initiatives, it is possible that the IMF can assist these countries through its lending facilities. Support for emerging markets and developing countries is an important priority of the IMF \cite{loayza2020macroeconomic}. Developed countries are still more economically unstable than industrialized economies and these are now especially hard hit by a shortage of medical supplies, a sudden impact on the world economy, capital flight, and, for some, a dramatic drop in commodity prices.

The European Union has implemented many data-driven strategies to face the formidable challenge brought by COVID-19. Despite the closure of borders and limitations of mobility, financial support has been provided to find effective vaccines, promote treatment therapies, protect salaries, etc. According to statistics from 2,061 adults, more than 82\% respondents think that the government should first consider the health and wellbeing of citizens. About 61\% of respondents are in favour of promoting social and environmental outcomes when the COVID-19 pandemic is over. Based on the report published by the Office for National Statistics, the economic growth will be significantly affected. A report entitled ``Tragedy of Growth" published by Positive Money on Monday points out that more attention should be paid on social and environmental indicators to save lives and improve the environment. A series of policies that can promote wellbeing without first increasing GDP are also released in this report, including cancellations or reductions of household debt. It is inferred that the macroeconomic shocks of the COVID-19 pandemic include financial stability and risk. With the emergence of economic adversities, designing and implementing innovative policies are necessary for a long-term view.

\section{Work Resumption}
\label{sec:5}

\begin{figure}[ht]
  \centering
  \includegraphics[width=0.48\textwidth]{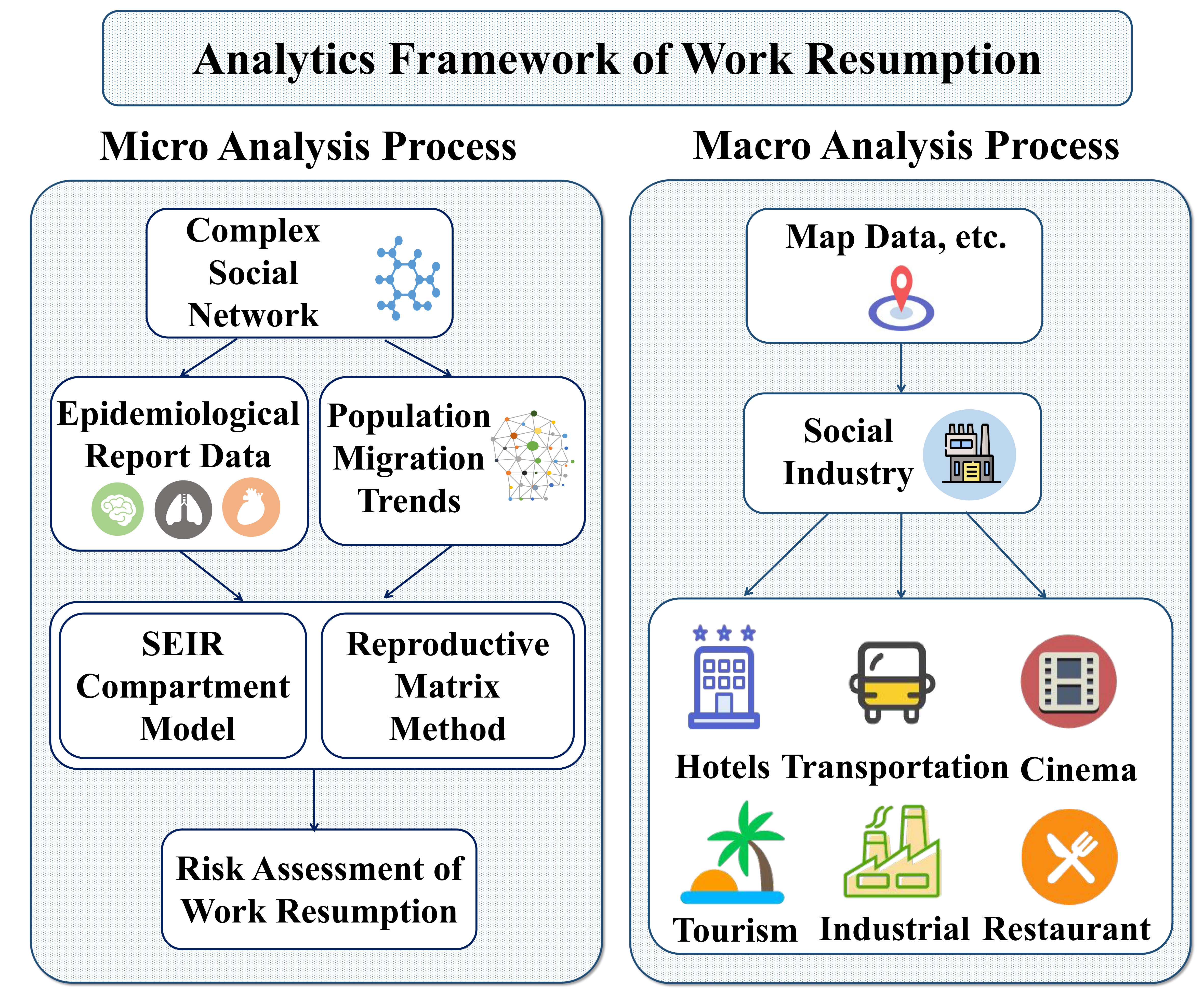}
  \caption{Data-driven analytics of work resumption.}
  \label{fig0}
\end{figure}

To handle the high infectivity of COVID-19, many countries have experienced several months of quarantine, social distancing, and travel restrictions. Although these measures can be effective, there is a corresponding influence upon  the economy. Governments in different countries must resume work to restore their economy affected by COVID-19~\cite{gentilini2020social,liu2020china}. However, work resumption is accompanied by the risk of epidemic rebound without social isolation, and the current long-term continuation of COVID-19 restrictions will lead to economic downturn and unpredictable social problems~\cite{mansilla2020manage}. Therefore, it is necessary to make appropriate decisions to balance the two contradictory forces of work resumption and epidemic rebound. In this section, we first discuss the recovery strategy in areas where the epidemic is under control. Then we introduce some related research on data-driven decision making for work resumption. Here, we give the analytics framework of work resumption from two perspectives, i.e., microscope and macroscope as shown in Fig.~\ref{fig0}.

Zhao et al.~\cite{zhao2020negligible} assessed the risks of COVID-19 rebounding caused by resumption of work. It was concluded that the probability of COVID-19 rebounding during resumption of work is very limited or even negligible. However, this work ignored the imported cases which are inconsistent with the current situation of COVID-19. Also, this work assumed that secondary attack rate of COVID-19 in enterprise clusters from an unidentified infected case is the same as that rate in family clusters, which is not correct for different countries~\cite{leung2020first,wang2020risk}.

Wang et al.~\cite{xia2020will} proposed a data-driven network modelling analysis of the work resumption. Epidemiological report data, and Baidu's population migration trends and distribution data are integrated to estimate the actual cumulative number of cases with the reproductive matrix method. The primary conclusions revealed that the risk of a second outbreak will be negligible under strong prevention and self-protection measures. This work aimed to assess the risk of recovery work from a micro perspective.

The work resumption of  is related to the economic impact of COVID-19. Some studies aim to quantify the impact of COVID-19 to provide decision making supports for work resumption from an indirect perspective. Baidu has used the Baidu map data to quantify the economic impact of COVID-19~\cite{huang2020quantifying}. The results show that travel-dependent industries (e.g. hotels, public transportation, etc.) have not yet recovered compared to same period in the past. Financial support should be provided to these sectors to strengthen the recovery. On the other hand, the sectors that are essential to human life such as workplaces, restaurants, and shopping venues have been recovering rapidly during the current period of work resumption.

Work resumption for migrant workers, which occupies a great proportion of the labour force, should also be noticed. Due to the social isolation and travel restrictions, these workers fail to return to urban areas to work. Furthermore, these workers usually exist in the industrial sectors such as export-oriented industry, hotels, tourism, and the entertainment industry. Many governments have taken specific measures to avoid unemployment for this group. For example, local governments have enhanced infrastructure building in rural areas to provide job opportunities for migrant workers. In addition, many countries attempt to start the progress of work resumption to recover their economies to some extent. Gentilini et al.~\cite{gentilini2020social} pointed out that labour market intervention is an effective way for the government to support work in the formal and informal sectors. Many countries, including Bosnia, Herzegovina, Romania, etc., have considered activation measures (worker training) to prepare for work resumption.

A community detection algorithm is utilized to identify local job markets in~\cite{bonato2020mobile}. The important conclusion reveals that the reduced mobility caused by social isolation and travel restrictions force the establishment of smaller local job markets, which has some impacts on the future work resumption. Barbieri et al.~\cite{barbieri2020italian} utilized the Italian Sample Survey on Professions data to assess the risk of infection of workers in 600 sectors. It showed that social isolation has effectively reduced the possibility of workers at risk of contagion. This work aims to provide data-driven decision support for policy makers who plan to adopt work resumption strategies.

Williams and Kayaoglu~\cite{williams2020covid} evaluated the undeclared worker group without access to financial support affected by COVID-19 and provided the possible recommendations to the government for the worker group in Europe. They analysed data from 27,565 interviews in 28 European countries and point out that the government policy needs adjustment during the period of COVID-19. Greater financial support to undeclared workers should be established by European governments. Bailey and West~\cite{bailey2020covid19} analysed the mortality and work resumption. Their analysis showed that the strict restrictions lead a total number of deaths at least eight times fewer than those who immediately resumed work. The conclusion may provide some suggestions while considering the work resumption strategies.

Rio-Chanona et al.~\cite{del2020supply} analysed the supply and demand shocks of the COVID-19 pandemic, which indicates that transportation and other industries are more vulnerable to demand-side shocks and entertainment. Aside from these impacts, hotels and tourism are facing constraints from both supply and demand. Furthermore, the study finds that the high-wage occupations are relatively immune from both supply-side and demand-side shocks while many low-wage occupations are much more economically vulnerable to these two shocks. For policymakers, this study suggests two important implications. The first is that it is vitally important to resume work as soon as possible without increasing the risk of epidemic rebound. The second is related to the issue of fairness emphasized in this study. The negative impact of COVID-19 on higher income knowledge and service workers can be negligible, while lower income workers receive greater consequence from this epidemic.

\section{Re-opening of Schools}
\label{sec:6}
\par Isolation measures can effectively reduce the infection rate and the speed of transmission. Therefore, policy makers in schools should first pay attention to non-pharmacological interventions, including delaying school attendance. Furthermore, the purpose of these measures is to establish social segregation and isolation. In the context of the epidemic, many studies focus on building models to analyse the impact of epidemic transmission and school closures. For example, Bayham and Fenichel \cite{bayham2020impact} studied the impact of school closures on the medical system.

However, the epidemiological benefits come with economic damage. Lempel et al. \cite{lempel2009economic} estimated the direct impact on the economy and health care of closing schools for 2, 4, 6, and 12 weeks. They discovered that the economic cost of closing all schools in for 4 weeks might be between $\$$$1$ billion and $\$$$47$ billion (0.1$\%$-0.3$\%$ of GDP), and it reduces primary healthcare personnel by 6$\%$ to 19$\%$. Therefore, after analyzing the school closures, when to reopen schools has become an important social issue.

\subsection{Re-opening Strategies}
\par In the context of the SARS outbreak, schools in the severely affected areas also chose to close. Normal teaching progress at many schools had been disrupted. The specific length of closure depended on the situation of different schools. After the spread of the SARS epidemic had weakened, the school re-opening strategies were different according to diverse circumstances.

\begin{figure*}[!t]
  \small
  \centering
  \includegraphics[width=\textwidth]{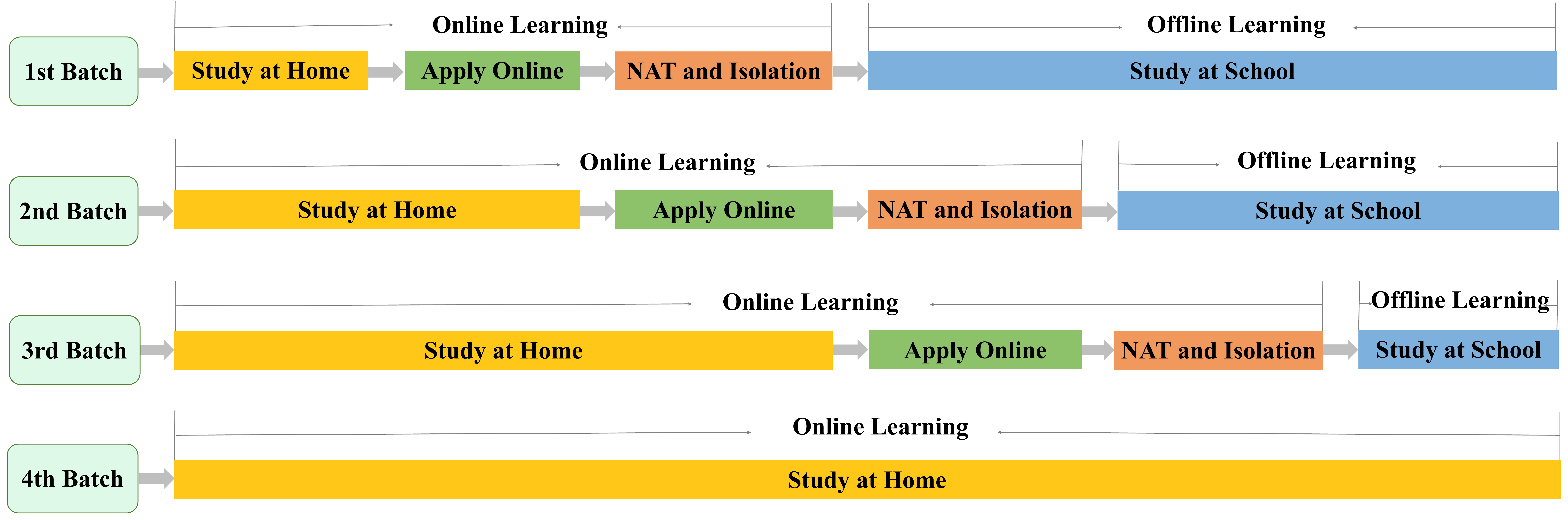}
  \caption{Student study managed in batches depending on the priority.} \label{fig1}
\end{figure*}

When the status of COVID-19 transmission in some areas is under control, schools should adopted a similar strategy to SARS. Specifically, for students who have urgent tasks and examinations, they are regarded as the first batch of students back to school. The rest of students are divided into different batches according to the relative level of urgency. For example, most local governments stipulate that the graduates of high schools are the first batch of students, university graduates are the second batch of students, primary school students are the third batch of students, and non-graduates from university are the fourth batch of students.

The time of re-opening school for each batch of students is reasonably divided. When the previous batch of students returns to school completely, the next batch of students is permitted to start returning to school. The entire process is roughly illustrated in Fig.~\ref{fig1}, in which nucleic acid test (NAT) and isolation may be required. Moreover, students can voluntarily choose whether to resume school. Significant recovery strategies guarantee schools not gather large numbers of students at the same time and avoid large numbers of people moving nationwide.

Wang et al. \cite{wang2020risk} proposed some solutions for, teaching, medical treatment, and other aspects. Singh and Adhikari \cite{singh2020age} calculated the basic reproduction rate $R_{0}$ and its time-related changes. On this basis, they study the impact of the duration of social alienation measures (i.e., closed schools). In the end, they concluded that a short period of social isolation after the outbreak cannot effectively prevent a second outbreak. They recommended a continuous lock-in program that relaxes regularly. Lee et al. \cite{lee2010simulating} simulated school activities and showed the significant effectiveness of closing schools. They found that it may be necessary to maintain some type of school suspension for at least 8 weeks to affect the overall serological seizure rate. They believed that the relatively short school suspension time (i.e. 2 weeks or less) can increase the overall seizure rate by returning susceptible students to school in the middle of the epidemic. Based on this, they concluded that individual and short suspensions may not quell the epidemic, but if they are maintained for at least 8 weeks, the spread of the epidemic can at least be delayed by one week. This provides more time to implement additional interventions and make existing interventions more effective.

\par For improvement of the decision making and decision support of re-opening schools, Duan et al. \cite{duan2013acp} used the largest H1N1 influenza outbreak as an example to establish an artificial society to simulate the epidemic in universities. They further proposed the Artificial societies, Computational experiments, and Parallel execution (ACP) method to more effectively control the outbreak of H1N1 influenza. The ACP model can be effectively applied in the improvement of intervention strategies. Moreover, in the application of strategy improvement, they concluded that the ACP method is useful for public health emergency management. In addition, Popa \cite{popa2020decision} found that there are no automated tools to suggest which decisions to make. Based on this, they first designed an algorithm that abstracts decisions into a combinatorial optimization problem. Finally, they show the integer linear program formula of the problem.
As for the decision support for the re-opening of school, Araz et al. \cite{araz2011simulation} modeled the spread of pandemic influenza in local universities and evaluate university mitigation policies. Beaton et al. \cite{beaton2007pandemic} provided a table-top exercise to evaluate the plans and policies used by the University of Washington (UW) to deal with influenza pandemics. This work reveals gaps in university pandemic influenza plans and policies. Issues analysed in this work include quarantine, decision making for re-entry, mental health services, and tracking the travel of personnel. Finally, they provide policy and planning suggestions on these issues. Based on the trade-off between the cost and benefit of the intervention strategy, Cao et al. \cite{cao2014evaluating} also analysed the various decisions of the epidemic situation, and their research can also be used to support the relevant decision making for re-opening of schools.

\subsection{Online Education}
\par A large number of schools have used the Internet for remote learning to ensure the continuity of teaching during the epidemic. In related research in educational psychology, multitasking has a negative impact on students' academic performance. This theory has been effectively proven by investigating the self-efficacy of self-regulated learning (SESRL) \cite{zuffiano2013academic}. But in online education, SESRL eases the contradictory relationship between multitasking and academic performance \cite{alghamdi2020online}. The moderated mediation effect of self-efficacy is only found in online classrooms. Based on the above findings, research in \cite{alghamdi2020online} shows that when students with high SESRL levels study online, the negative impact of multitasking on the score is reduced. As for the public acceptance of online education, Samad and Khalid\cite{samad2019acceptance} survey 125 elementary school science teachers using a 10-item questionnaire and analyse the questionnaire data obtained. Their research results show that primary school science teachers have a high acceptance of using online learning. And based on an online survey of 607 valid responses gathered from 896 online respondents, Vate \cite{vate2020psychological} discovered that there is a significant positive correlation between students' attitudes towards online education and students' life satisfaction. However, there are still many problems in online education, and these problems pose challenges to the quality of learning, for example, academic integrity issues. In response to these issues, Ohio State University \cite{bane2019academic} has adopted a dual approach to ensure the academic integrity of its online education products. However, to meet the needs of the changing educational situation, a complete solution is still needed.

\section{Challenges and Open Issues}
\label{sec:7}

Many strict policies have been taken to contain the COVID-19 pandemic, whereas the impact of the enforcement and subsequent loosening of these policies have not been very well analysed or understood. Assumptions, premises, and the context are necessary in the decision making process. All of these are required to be verified through data in data-driven policymaking. Data-driven research can better guide policymaking if the data are appropriately analysed. It is generally believed that data-based decision making will lead to better decisions compared to decisions based on observation or informed guesswork. However, data-driven policy making during COVID-19 is quite special. For one thing, traditional data-driven policy making is extrapolated from key data sets that contain historical data. Nevertheless, we have no historical data for COVID-19, which will definitely restrain the ability and adaptability of decision making.

How can we best make decisions with regard to available data? It appears that consistent, comparable, and traceable data are becoming more important under these circumstances. Various heterogeneous data sources can provide abundant data but bring about privacy problems at the same time. Further study is needed about how to mine these data and develop efficient policies that can meanwhile protect public privacy as much as possible. Also, how to evaluate and balance the importance of different measurements in relation to data-driven policy making. Since not enough historical data can be employed, there is the possibility of using intuition-driven decisions. Here, we list three fundamental challenges and open issues that are considered important to tackle first.

\textbf{Data Collection and Quality: }Effective data collection and high-quality data are decisive to data processing. Data-driven decision making is based on both well-collected and well-processed data. The quality of data is extremely significant, especially when data is used to guide decision making such as financial aid, resumption of work or school, etc. Data quality is generally evaluated by accuracy, precision, correctness, timeliness, etc. It is significantly difficult to collect data and at the same time ensure the quality of data. Current methods and tools can be applied to collect COVID-19 data, but COVID-19 data will also need some specialized methods to process the data to ensure quality. Therefore, further work should focus on how to collect COVID-19 data effectively and at the same time ensure the quality of data.

\textbf{Complex Data Analysis: }The rapid spread of COVID-19 has brought about diverse and abundant data, which are represented in nearly all areas of life. Under these exceptional circumstances, all the relevant data should be integrated to guide COVID-19 decision making. However, COVID-19 related data include basic information, transportation data, diagnosis data, scientific research data, and other kinds of data from multiple-sources, which makes it especially difficult to process. High dynamism, heterogeneity, as well as unpredictability are also properties of COVID-19 data. Analyzing such data requires both hardware equipment and highly efficient algorithms, and currently there are no such specialized ones. Thus, further studies should pay additional attention to data analysis of highly dynamic, heterogeneous, multi-sources, as well as unpredictable sources, including new processing methods and analytic tools.

\textbf{Fairness of Decision Making: }Fairness is always one of the most important problems in decision making. For some time, decision making has been believed to be more reliable when data driven rather than human based. However, even well designed and correctly implemented algorithms may still make decisions with prejudice. Such prejudice can be reduced by improving the fairness and interpretability of machine learning algorithms. It is generally acknowledged that the judgement of humans is also an important perspective to ensure fairness of decision making. Therefore, there is a need for more research when implementing data-driven decision making generally, and not just limited to the COVID-19 response.

\section{Conclusion}
\label{sec:8}
Data-driven decision making has been demonstrated to be both important and effective for the COVID-19 response. Different countries and regions have implemented many policies that are data-driven, such as prevention and control policies, psychological counselling policies, financial aid policies, and resumption/re-opening policies. In this paper we have summarized these policies including examples from all around the world. From different kinds of data related to COVID-19, policy makers have developed large volumes of useful information, and based on this information and machine learning algorithms, data are guiding us to better and more appropriate choices. We first discussed COVID-19 related data and some prevention and control policies and then described the psychological counselling policies driven by COVID-19 data. In the financial aid section, we have discussed the protection of lives as well as economic recovery. Following on, we introduced in detail policies in the resumption of both work and school. Finally, we listed the current challenges and open issues, including data collection and quality, complex data analysis, and fairness of decision making.






\bibliographystyle{ACM-Reference-Format}
\bibliography{reference}


\end{document}